# Infant Cry Emotion Recognition Using Improved ECAPA-TDNN with Multi-scale Feature Fusion and Attention Enhancement

*Junyu Zhou, Yanxiong Li\*, Haolin Yu*

School of Electronic and Information Engineering, South China University of Technology, Guangzhou, China
202230324417@mail.scut.edu.cn, eeyxli@scut.edu.cn, 1430876441@qq.com

## Abstract

Infant cry emotion recognition is crucial for parenting and medical applications. It faces many challenges, such as subtle emotional variations, noise interference, and limited data. The existing methods lack the ability to effectively integrate multi-scale features and temporal-frequency relationships. In this study, we propose a method for infant cry emotion recognition using an improved Emphasized Channel Attention, Propagation and Aggregation in Time Delay Neural Network (ECAPA-TDNN) with both multi-scale feature fusion and attention enhancement. Experiments on a public dataset show that the proposed method achieves accuracy of 82.20%, number of parameters of 1.43 MB and FLOPs of 0.32 Giga. Moreover, our method has advantage over the baseline methods in terms of accuracy. The code is at https://github.com/kkpretend/IETMA.

## 1. Introduction

Infant cries encode biologically significant information through acoustic features like frequency, duration, and timbre, correlating with emotional/physical states and influenced by individual variability and environmental noise. Dunstan's theory identifies primal, reflexive vocalizations in infants aged 0-3 months linked to specific physiological needs, verified in later studies [1]. Post-neonatal stages involve complex cries. Practical applications require multimodal integration of cry acoustics, behavioral cues, and contextual data.

Recent advances in signal processing and data-driven algorithms have enabled infant cry analysis for detection [2]-[5], cause classification [6]-[13], and pathological diagnosis [14]-[16]. Traditional methods rely on handcrafted acoustic features (e.g., Mel-frequency cepstral coefficients, linear prediction coefficient, spectral centroid) combined with classifiers (e.g., support vector machine, random forest, logistic regression [2], [6]-[8]. Hybrid methods, such as multistage ensemble models, achieve accuracies up to 93.7% [8].

Deep-learning based methods exceed the above traditional methods by extracting deep embeddings from samples. Transfer learning with pretrained Convolutional Neural Network (CNNs) or spectrogram-based network of long-short term memory attains accuracy of 92~99% for cry detection [5] and cry classification [9]. CNNs struggle with long-term dependencies, whereas recurrent neural networks (RNNs) are computationally inefficient. Models combining CNNs, RNNs, and attention mechanisms can further improve the performance of emotion recognition [10]-[12].

Desplanques et al. propose the ECAPA-TDNN with an attention module of Squeeze Excitation (SE), and win first place at the VoxSRC2020 competition [17]. In a recent study [18], the ECAPA-TDNN is used for speaker verification based on infant cry, and an equal error rate of 25.8% is obtained. It is shown that the ECAPA-TDNN can learn effective embeddings from samples, and obtain higher accuracy than other models.

Infant cries are emotionally nuanced, whose characteristics are hard to be captured by the model. Environmental noises further disrupt infant cries, and thus degrade the model's performance. In addition, training datasets of infant cries are relatively small-scale, which leads to a decrease in the model's generalization ability and adaptability. The ECAPA-TDNN has three critical limitations for emotion recognition of infant cries.

First, its single-scale convolutional design with fixed kernel sizes and fixed dilation rates restricts the receptive field adaptability. As a result, multi-scale emotional information cannot be effectively captured. For instance, pain-induced high-frequency spikes require small receptive fields for precise localization, while hunger-induced rhythmic patterns demand larger scales to discern trends. Hence, single-scale convolution will result in blurry information and incomplete utilization of multi-scale emotional information.

Second, the ECAPA-TDNN lacks effective interlayer feature fusion mechanisms. Shallow layers capture low-level acoustic features, while deep layers extract abstract representations. The absence of cross-layer integration hampers the comprehensive interaction of features.

Third, the temporal and channel attention mechanisms operate in isolation without synergistic interaction. Temporal attention focuses on time-series importance, while channel attention emphasizes frequency-band significance. However, the isolated usage of temporal and channel attentions prevents the model from effectively integrating the time-frequency intertwined emotional cues. The disconnection between temporal and channel attention modules limits the model's holistic feature integration capability and decision-making capability, and finally reduces the model's recognition accuracy.

In short, the limitations in scale adaptability, hierarchical feature fusion, and attention synergy constrain the ECAPA-TDNN's performance in recognizing infant cry emotion. To overcome the above deficiencies of the ECAPA-TDNN, we propose an improved ECAPA-TDNN for infant cry emotion recognition in this paper.

## 2. Method

### 2.1 Method Framework

The framework of the improved ECAPA-TDNN is the same as that of the ECAPA-TDNN, except for three synergistic modules. Figure 1 illustrates the frameworks of both ECAPA-TDNN and improved ECAPA-TDNN.

In the improved ECAPA-TDNN, the modules of Residual Squeeze-and-Excitation (RSE), Multi-scale Channel Attention (MCA) and differential attention are incorporated into the architecture of the ECAPA-TDNN. In addition, the global average pooling and Softmax substitute for the attentive statistics pooling and AAM-Softmax of the ECAPA-TDNN, respectively. The RSE module implements cross-dimensional interaction between the time-varying patterns and critical frequency bands, and thus effectively bridges the temporal and spectral domains. The MCA module enhances the feature extraction backbone by parallel multi-scale dilated convolution, and thus captures both localized spectro-temporal structures and extended contextual dynamics in the infant vocalizations. The global average pooling compresses the channel-wise feature representations, followed by the differential attention module, which can correct the fused channel weights by emphasizing emotion-relevant channels based on learned interdependencies. In summary, the layers within the blue dotted-line box in Figure 1 (b) not only enhance the interaction of features between layers but also prevent the emergence of issues such as gradient vanishing.

\* Corresponding author: Yanxiong Li (eeyxli@scut.edu.cn)

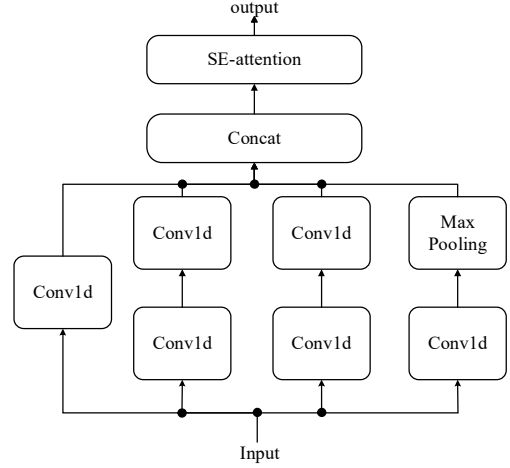

Figure 2 *Structure of MCA convolution layer.*

and dilation rate $d_k$. The dilated convolutions can be used to obtain exponential expansion of receptive fields without increasing computational complexity. The multi-scale features are then concatenated along the channel dimension, namely

$$F_{cat} = [F_3, F_5, F_7, M] \in \mathbb{R}^{(4C \times T)}, \quad (2)$$

where $F_3$, $F_5$ and $F_7$ are calculated by Equation (1), and $M$ is defined by

$$M = MaxPool(Conv_{1 \times 1}(X)) \in \mathbb{R}^{(H \times W \times C)}. \quad (3)$$

$$z = (1/T) \sum_{t=1}^{T} F_{cat}(t) \in \mathbb{R}^{4C}, \quad (4)$$

where $T$ denotes the number of time steps. Subsequently, channel dependencies are learned through a gating mechanism using two fully-connected layers:

$$s = \sigma(W_2 \delta(W_1 z)) \in \mathbb{R}^{3C}, \quad (5)$$

where $\delta(\cdot)$ and $\sigma(\cdot)$ denote ReLU function and Sigmoid activation function, respectively; $W_1 \in \mathbb{R}^{(3C/r) \times 3C}$ and $W_2 \in \mathbb{R}^{3C \times (3C/r)}$ represent trainable weight matrices with reduction ratio $r$. The output of the block is obtained through channel-wise scaling:

$$F_{out} = s \odot F_{cat} \in \mathbb{R}^{4C \times T} \quad (6)$$

where $\odot$ denotes channel-wise multiplication. The original spatial resolution can be preserved, and discriminative channel responses can be enhanced. This approach for multi-scale feature extraction, coupled with the SE mechanism to adjust channel weights, not only addresses the issue of ECAPA-TDNN's single-scale feature extraction but also maintains the coherence of the attention mechanism. This is crucial for the classification of infant crying causes, because the usage of global pooling operations avoids the premature reduction of dimensions, which will lead to the neglect of the importance of feature extraction.

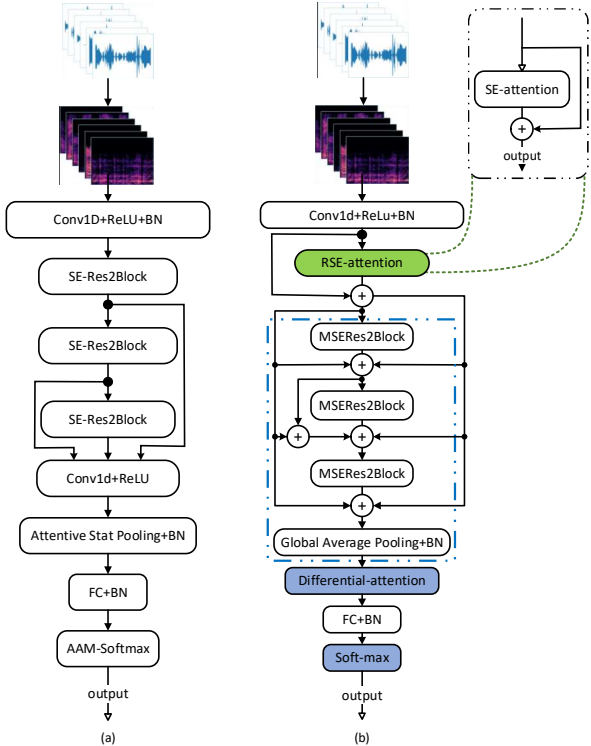

Figure 1 *The frameworks of (a) ECAPA-TDNN and (b) improved ECAPA-TDNN. As shown in subgraph (b), the improved parts include modules in the green background box, modules in the blue background box, and modules in the blue dotted-line box.*

### 2.2 Multi-Scale Channel Attention

To enhance the model's ability to extract discriminative features across different receptive fields, we propose a MCA module that integrates multi-scale feature extraction with channel-wise adaptive weighting [13], [19]. As illustrated in Figure 2, the MCA module consists of two key components: parallel convolutional branches with varying dilation rates for multi-scale context aggregation, and a channel attention mechanism for feature recalibration.

For an input feature map $X \in \mathbb{R}^{(C \times T)}$, the multi-scale transformation is defined by

$$F_k = Conv_{1 \times k}^{(d=d_k)}(Conv_{1 \times 1}(X)) \quad (1)$$

where $Conv_{1 \times k}^{(d=d_k)}$ denotes a convolution with kernel size $1 \times k$

### 2.3 Temporal-Channel Interactive Attention Mechanism

The temporal-channel interactive attention module performs information interaction between temporal dimension and channel dimension via a RSE structure. Given an input feature $X \in \mathbb{R}^{(T \times C)}$, where $T$ and $C$ denote numbers of time steps and channels, respectively. The processing pipeline is formalized by

$$X_{rse} = RSE(X) = X + SE(X), \quad (7)$$

where $X_{rse}$ represents the features after the merging of time and frequency domains, and $SE(\cdot)$ denotes the Squeeze-and-Excitation operation, and is defined by

$$SE(X) = \sigma(W_2 \delta(W_1 z)) \odot X, \quad (8)$$

with
$$z = (1/T) \sum_{t=1}^{T} X(t) \in \mathbb{R}^C, \quad (9)$$

where $W_1 \in \mathbb{R}^{(C/r \times C)}$ and $W_2 \in \mathbb{R}^{(C \times C/r)}$ represent learnable parameters; $\sigma(\cdot)$, $\odot$ and $T$ stand for a Sigmoid function, channel-wise multiplication and the number of time steps, respectively. The RSE module preserves original temporal resolution and enhances channel adaptability. Temporal attention output $A_t \in \mathbb{R}^{(T \times 1)}$ and channel attention output $A_c \in \mathbb{R}^{(1 \times C)}$ are fused by

$$M = (A_t A_c^T) \odot X_{rse}. \quad (10)$$

The final output integrates original temporal features by

$$X_{out} = X + Conv1D(M). \quad (11)$$

For pain-related infant cries (e.g., 0.8-1.2 kHz abrupt onsets), temporal attention peaks at onset t_0 induce elevated weights for 0.8-1.2 kHz channels via $A_t A_c^T$ correlation.

The MCA and RSE Res2Block incorporates the RSE and MCA mechanisms into the original Res2Block of the ECAPA-TDNN. Its structure is illustrated in Figure 3.

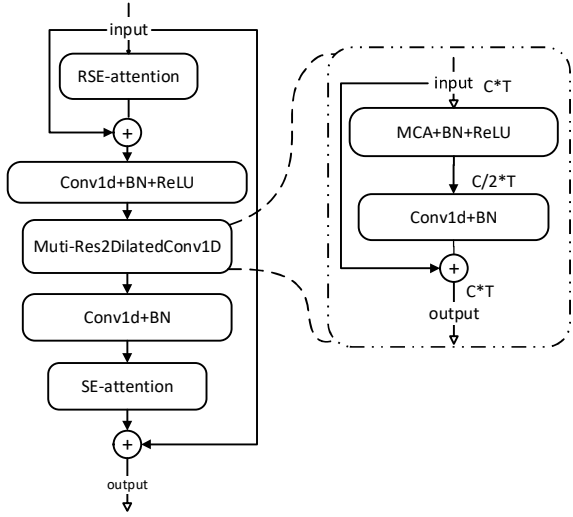

Figure 3 *Structure of MCA and RSE Res2Block.*

As illustrated in Figure 3, after passing through the RSE-attention module, the features are further processed by a one-dimensional convolution and a Muti-Res2DilatedConv1D module. Afterwards, the channel weights are adjusted via the SE-attention module.

**2.4 Channel Recalibration via Differential Attention**

Like global average pooling and multi-scale feature fusion, the concatenated feature tensor $Z \in \mathbb{R}^{(N \times C)}$ undergoes channel-wise recalibration by a differential attention operation of

$$Z_{out} = (A_1 - \lambda A_2) V \times Z \quad (12)$$

where:

$$A_1 = Softmax(Q_1 K_1^T / \sqrt{d}) \quad (13)$$

$$\begin{cases} Q_1 = Split(Z W_{Q_{z1}}) \\ K_1 = Split(Z W_{K_1}) \end{cases} \quad (14)$$

$$A_2 = Softmax(Q_2 K_2^T / \sqrt{d}) \quad (15)$$

$$\begin{cases} Q_2 = Split(Z W_{Q_2}) \\ K_2 = Split(Z W_{K_2}) \end{cases} \quad (16)$$

$$V = Z W^V \in \mathbb{R}^{(N \times C)} \quad (17)$$

where $\lambda \in \mathbb{R}$ is a learnable scalar optimized via gradient descent; T denotes the transpose of matrix; and Split(·) is an operation for dividing a matrix into smaller matrices along a specified dimension. Each smaller matrix is then multiplied by corresponding weight matrices $W_Q$ and $W_K$ to create multiple query and key matrices.

The dual Softmax subtraction ($A_1 - \lambda A_2$) implements contrastive channel weighting, suppressing attention noise and amplifying discriminative spectral patterns [20], [21]. This is achieved by adaptively canceling attention scores shared across both pathways ($A_1$ and $A_2$), which correspond to non-informative features. In the fused channels of baby crying, different channels contribute to emotional expression to varying degrees. The incorporation of this module further enhances the model's ability to capture information from specific channels.

## 3. Experiments

**3.1 Experimental Datasets**

The experiments are conducted on a public dataset which is released by iFLYTEK and the University of Science and Technology of China in 2020. This dataset contains a total of 3000 audio clips. The duration of each audio clip ranges from 3 to 5 seconds. There are six types of infant cry emotions in this dataset, namely "awake", "diaper", "hug", "hungry", "sleepy", and "uncomfortable". The ratio of training set to testing set is equal to 8:2. The detailed information of this dataset can be found at: https://aistudio.baidu.com/datasetdetail/41960.

**3.2 Experimental Setup**

The hardware for performing the experiments mainly consists of an NVIDIA GeForce RTX 3060 GPU. The software mainly includes Python 3.10 and the deep learning framework of PyTorch 2.4. The audio clips are preprocessed by removing silent segments and normalizing the lengths to ensure uniformity. The model is trained for 700 epochs using the Adam optimizer with a batch size of 64, and the learning rate is set to $2^{-5}$. Additionally, 13 filters are adopted for spectral feature extraction, while 128 channels are used in the neural network.

To verify the effectiveness of the improved ECAPA-TDNN, some necessary experiments are conducted. First, we conduct ablation experiments to show the performance contributions of different key modules of the improved ECAPA-TDNN. Then, we compare the improved ECAPA-TDNN with the ECAPA-TDNN and Resnet18 using multiple performance metrics. The performance metrics include accuracy, number of model parameters (Param), and Floating-Point Operations (FLOPs) required for inference per sample. In addition, the confusion matrices of different models are presented for illustrating the details of confusions between different types of emotions obtained by different models.

**3.3 Experimental Results**

Table 1 presents the ablation experimental results obtained by the improved ECAPA-TDNN with different combinations of MCA, RSE and differential-attention modules. From the results of Table 1, we can draw two conclusions. First, each of the above three modules has an impact on the accuracy obtained by the proposed model for infant crying emotion recognition. Moreover, the MCA module has the greatest impact on the accuracy, since the accuracy drops from 82.20% to 78.37% after removing it. Second, when all the three modules are used together, the proposed model obtains the highest accuracy score of 82.20%. Hence, each module has complementarity and can work together to improve the accuracy of the proposed model.

Table 1 *Ablation experimental results.*

| Model | Accuracy |
|---|---|
| Improved ECAPA-TDNN with all modules | 82.20% |
| Improved ECAPA-TDNN w/o MCA | 78.37% |
| Improved ECAPA-TDNN w/o RSE | 78.54% |
| Improved ECAPA-TDNN w/o differential-attention | 78.70% |

The improved ECAPA-TDNN is compared with two state-of-the-art models, namely the ECAPA-TDNN and the Resnet18. The results obtained by different models are listed in Table 2. The improved ECAPA-TDNN obtains the accuracy score of 82.20%. Its accuracy score is higher than that obtained by both ECAPA-TDNN and Resnet18. The accuracy gain obtained by the improved ECAPA-TDNN is mainly attributed to the integration of multi-scale feature extraction, optimization of the connection and coherence of the attention mechanism. The multi-scale feature extraction module can capture rich feature representations of the samples. The cross-layer connection module can realize the comprehensive interaction of features. The interaction between temporal and channel attention modules can improve the model's abilities of holistic feature integration and decision-making. Hence, the improved ECAPA-TDNN achieves higher accuracy scores compared to the two state-of-the-art models, namely ECAPA-TDNN and Resnet18.

Table 2 *The results obtained by different models.*

| Model | Accuracy (%) | Param (MB) | FLOPs (Giga) |
|---|---|---|---|
| Improved ECAPA-TDNN | 82.20 | 1.43 | 0.32 |
| ECAPA-TDNN | 73.38 | 0.84 | 0.20 |
| Resnet18 | 60.07 | 10.66 | 10.47 |

The number of parameters and the FLOPs of the improved ECAPA-TDNN are 1.43 MB and 0.32 Giga, respectively. These two values are higher than the counterparts of the ECAPA-TDNN but lower than the counterparts of the Resnet18. That is, it doesn't and does have advantage over the ECAPA-TDNN and the Resnet18 in complexity, respectively.

The confusion matrix of the improved ECAPA-TDNN for classifying infant cry emotions is shown in Figure 5. There are differences in the degree of confusion between different categories. The confusion between "awake" and "hungry" is the highest (reaching 12.15%). However, the confusion between "sleepy" and "diaper", and the confusion between "awake" and "uncomfortable", reach the lowest value of 0.

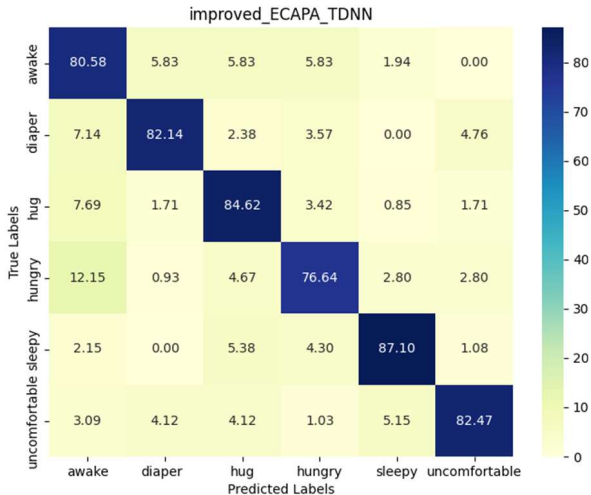

Figure 5 *Confusion matrix of the improved ECAPA-TDNN.*

For comparison convenience, Figures 6 and 7 depict the confusion matrices of the ECAPA-TDNN and Resnet18, respectively. From the results in Figure 5 to Figure 7, it can be known that the ECAPA-TDNN and Resnet18 have higher levels of confusion for various categories of emotions than the improved ECAPA-TDNN.

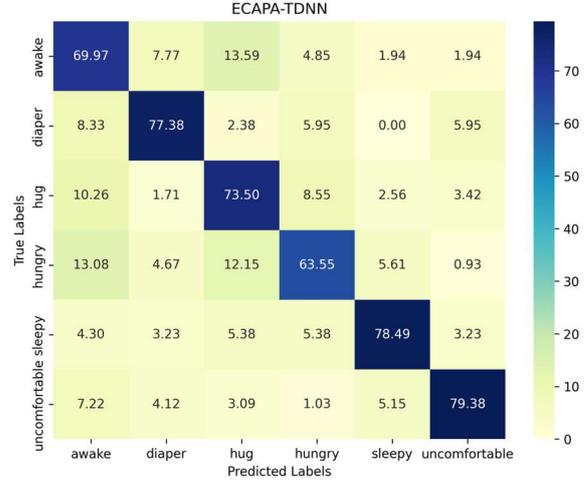

Figure 6 *Confusion matrix of the ECAPA-TDNN.*

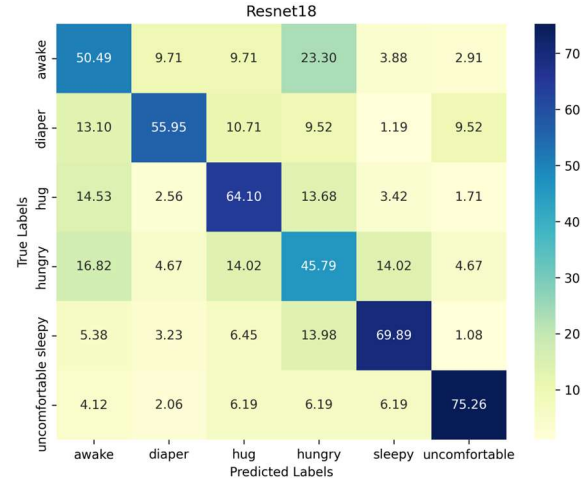

Figure 7 *Confusion matrix of the improved Resnet18.*

## 4. Conclusions

In this paper, we design a model of the improved ECAPA-TDNN for infant cry emotion recognition. It integrates multi-scale feature fusion module and bidirectional temporal-channel co-attention modules. Experimental results demonstrate that the proposed model achieves the accuracy score of 82.20% which is higher than that obtained by two state-of-the-art models, namely ECAPA-TDNN and Restnet18. Hence, it has advantage over other two models in accuracy.

However, the proposed model still has its own limitations. First, its complexity is higher than the ECAPA-TDNN. Second, its attention modules mainly concentrate on the temporal-channel interactions, but spatial dependencies in spectrogram representations remain underutilized. In the future work, we will explore hybrid attention architectures by incorporating spatial-spectral relationships. We will also take measures to reduce the complexity of the model for deploying it on edge devices, such as knowledge distillation, pruning, and parameter compression of models.

## 5. Acknowledgement

This work was supported by the national natural science foundation of China (62371195, 62111530145, 61771200), the exchange project of the 10th Meeting of the China-Croatia Science and Technology Cooperation Committee (10-34), the provincial undergraduate training program for innovation and entrepreneurship (S202410561216), and Guangdong provincial key laboratory of human digital twin (2022B1212010004).

## 6. References

[1] C. A. Bratan, M. Gheorghe, I. Ispas, E. Franti, M. Dascalu, S. M. Stoicescu, I. Roşca, F. Gherghiceanu, D. Dumitrache, and L. Nastase, "Dunstan baby language classification with CNN," in *Proc. of International Conference on Speech Technology and Human-Computer Dialogue (SpeD)*, Bucharest, Romania, 2021, pp. 167-171.

[2] X. Yao, M. Micheletti, M. Johnson, E. Thomaz, and K. de Barbaro, "Infant crying detection in real-world environments," in *Proc. of IEEE International Conference on Acoustics, Speech and Signal Processing*, Singapore, 2022, pp. 131-135.

[3] T. Khandelwal, R. K. Das, and E. S. Chng, "Is your baby fine at home? Baby cry sound detection in domestic environments," in *Proc. of Asia-Pacific Signal and Information Processing Association Annual Summit and Conference (APSIPA ASC)*, Chiang Mai, Thailand, 2022, pp. 275-280.

[4] L. S. Foo, W.-S. Yap, Y. C. Hum, Z. Kadim, H. W. Hon, and Y. K. Tee, "Real-time baby crying detection in the noisy everyday environment," in *Proc. of IEEE Control and System Graduate Research Colloquium (ICSGRC)*, Shah Alam, Malaysia, 2020, pp. 26-31.

[5] S. P. Narayanan, M. S. Manikandan, and L. R. Cenkeramaddi, "Spectrogram and LSTM based infant cry detection method for infant wellness monitoring systems," in *Proc. of International Conference on Human System Interaction (HSI)*, Paris, France, 2024, pp. 1-4.

[6] G. Aggarwal, K. Jhajharia, J. Izhar, M. Kumar, and L. Abualigah, "A machine learning approach to classify biomedical acoustic features for baby cries," *Journal of Voice*, vol. 38, no. 2, pp. 245-253, 2023.

[7] L. Liu, W. Li, X. Wu, and B. X. Zhou, "Infant cry language analysis and recognition: an experimental approach," *IEEE/CAA Journal of Automatica Sinica*, vol. 6, no. 3, pp. 778-788, May 2019.

[8] V.R. Joshi, K. Srinivasan, P.M.D.R. Vincent, V. Rajinikanth, and C.Y. Chang, "A multistage heterogeneous stacking ensemble model for augmented infant cry classification," *Frontiers in Public Health*, vol. 10, p. 819865, Mar. 2022.

[9] G. Anjali, S. Sanjeev, A. Mounika, G. Suhas, G. P. Reddy, and Y. Kshiraja, "Infant cry classification using transfer learning," in *Proc. of IEEE Region 10 Conference on TENCON*, Hong Kong, 2022, pp. 1-7,

[10] X. Shen, B. Lv, T. Liu, and Q. Cheng, "Infant speech emotion recognition based on channel attention mechanism with ResNet-BiLSTM," in *Proc. of International Conference on Information Science, Parallel and Distributed Systems (ISPDS)*, Guangzhou, China, 2024, pp. 54-57.

[11] S. G. A, G. S, G. Tharagarani, S.P, and S. B, "An automated mood analysis of crying infants through sound recognition using hybrid deep learning," in *Proc. of International Conference on Smart Technologies and Systems for Next Generation Computing (ICSTSN)*, Villupuram, India, 2024, pp. 1-6.

[12] Y. Liu, B. Lv, S. Xu, and X. Shen, "Emotion recognition of infant crying sounds using convolutional recurrent neural network with multi-scale joint attention mechanism," in *Proc. of International Conference on Information Systems Engineering (ICISE)*, Dalian, China, 2023, pp. 615-619.

[13] J. Yang, Z. Zhang, J. Li, and C. Lin, "A multi-scale convolutional attention neural network based on residual block downsampling for infant cry classification and detection," in *Proc. of International Conference on Internet of Things, Automation and Artificial Intelligence (IoTAAI)*, Guangzhou, China, 2024, pp. 9-14.

[14] C. Ji, T. B. Mudiyanselage, Y. Gao, and Y. Pan, "A review of infant cry analysis and classification," *EURASIP Journal on Audio, Speech, and Music Processing*, vol. 2021, no. 1, Art. no. 8, 2021.

[15] J. J. Parga, S. K. Lewis, and M. H. Goldstein, "Defining and distinguishing infant behavioral states using acoustic cry analysis: Is colic painful ?," *Pediatric Research*, vol. 87, no. 3, pp. 440-447, 2020.

[16] A. Gorin, C. Subakan, S. Abdoli, J. Wang, S. Latremouille, and C. C. Onu, "Self-supervised learning for infant cry analysis," in *Proc. of ICASSP Workshop on Safety and Security in Speech and Biomedical Signal Processing (SASB)*, IEEE, 2023, pp. 1-5.

[17] B. Desplanques, J. Thienpondt, and K. Demuynck, "ECAPA-TDNN: Emphasized channel attention, propagation and aggregation in TDNN based speaker verification," in *Proc. of INTERSPEECH*, 2020, pp. 3830-3834.

[18] D. Budaghyan, C. C. Onu, A. Gorin, C. Subakan, and D. Precup, "CryCeleb: A speaker verification dataset based on infant cry sounds," in *Proc. of IEEE International Conference on Acoustics, Speech and Signal Processing*, Seoul, Korea, 2024, pp. 11966-11970.

[19] S. Dixit, A. Jain, and R. Singh, "Improving speaker representations using contrastive losses on multi-scale features," *arXiv preprint arXiv:2410.05037*, 2024.

[20] T. Ye, L. Wang, and H. Li, "Differential transformer," *arXiv preprint arXiv:2410.05258*, 2024.

[21] F. Li, C. Cui, and Y. Hu, "Classification of Infant Crying Sounds Using SE-ResNet-Transformer," *Sensors*, vol. 24, no. 20, pp. 6575, 2024.